\documentclass[preprint,12pt]{elsarticle}

 \usepackage{graphicx}
 \usepackage{float}

\usepackage{amssymb}

\journal{Journal of Nuclear Materials}

\begin{document}

\begin{frontmatter}



\title{Surface Morphology and Phase Stability of Titanium Foils
Irradiated by 136 MeV $^{136}$Xe
}


\author[label1]{S. Sadi}
\author[label1]{ A. Paulenova}
\author[label2]{W. Loveland\corref{cor1}\fnref{fn1}}
\author[label2]{P.R. Watson}
\author[label3]{J. P. Greene}
\author[label3]{ S. Zhu}
\author[label3]{G. Zinkann}

\address[label1]{Dept. of Nuclear Engineering and Radiation Health Physics,  Oregon State University, Corvallis, OR 97331, USA}
\address[label2]{Dept. of Chemistry, Oregon State University, Corvallis, OR 97331, USA}
\address[label3]{Physics Division, Argonne National Laboratory, Argonne, IL 60439, USA}

\cortext[cor1]{Corresponding Author}
\fntext[fn1]{Email: lovelanw@onid.orst.edu (W. Loveland).}

\begin{abstract}

A stack of titanium foils was irradiated with 136 MeV $^{136}$Xe to study microstructure damage and phase stability of titanium upon irradiation. X-ray diffraction, scanning  electron microscopy/energy dispersive spectroscopy  and atomic force microscopy were used to study the resulting microstructure damage and phase stability of titanium. We observed the phase transformation of polycrystalline titanium  from alpha-Ti (hexagonally closed packed (hcp)) to face centered cubic (fcc)  after irradiation with 2.2 x 10$^{15}$ ions/cm$^{2}$.  Irradiation of Ti with  1.8 x 10$^{14}$-2.2 x 10$^{15}$ ions/cm$^{2}$ resulted in the formation of voids, hillocks, dislocation loops, dislocation lines, as well as polygonal ridge networks. 

\end{abstract}

\begin{keyword}
Microstructure damage, phase stability, titanium,  X-ray 
                   diffraction, SEM/EDS, and AFM

\end{keyword}

\end{frontmatter}


\section{Introduction}
\label{}

Microstructure changes and phase transformations in titanium due to radiation damage are of great interest in nuclear science and engineering.  Interest in the investigation of this material at elevated temperatures is motivated by its use in fusion reactors and the Reduced Enrichment Research and Test Reactor (RERTR) program. 

Titanium undergoes an allotropic phase transformation at 882.5 $^{\circ}$C, changing from a hexagonally closed-packed crystal structure (hcp) ($\alpha$-phase) to a body-centered cubic (bcc) crystal structure ($\beta$-phase). Ti also transforms to a simple hexagonal structure ($\omega$ phase) at high pressures \cite{Dammak,Lutjering}.  Also, the face centered cubic (fcc) -Ti phase has been found at ambient temperatures in thin metallic titanium.  A phase transformation from hcp to fcc titanium has been observed during irradiation by Ar ions. \cite{Josell,  Zhang}. Irradiation of titanium with 2.2 GeV U ions at 20K  (1.2 x 10$^{13}$ ions/cm$^{2}$) \cite{Dammak} resulted in a $\alpha$ $\to$ $\omega$ phase transformation.

There have been numerous attempts to investigate microstructure damage in titanium irradiated with swift heavy ions. Budzynski and co-workers \cite{Budzynski} investigated the microstructure surface damage of titanium  irradiated by 240-MeV Kr ions at a fluence of 1.0 x 10$^{14}$ ions/cm$^{2}$.  They also irradiated Ti with 130-MeV Xe ions at fluences of 1.0 x 10$^{14}$ - 4.9 x 10$^{14}$ ions/cm$^{2}$. They found that the most characteristic irradiated Ti surface features after irradiation by Kr and Xe ions are hillocks and craters(voids), respectively. Grain size was decreased by increasing the fluence of Kr ions and Xe ions; meanwhile, micro-mechanical properties such as micro-hardness and micro-strain increased after Kr and Xe ion irradiations.

The objective of this study was to study the crystallographic phase stability and microstructure damage behavior in Ti surfaces with high temperature irradiations using 136 MeV $^{136}$Xe ions with  fluences of 1.8 x 10$^{14}$ - 2.2 x 10$^{15}$ ions/cm$^{2}$.

\section{Experimental Methods}
\subsection{Sample Preparation}

Titanium metal foils were purchased from Goodfellow (TI000090) as pre-cut 5x5 cm squares with a thickness of 0.904 mg/cm$^{2}$ Ti  (0.002 mm, 99.6$\%$ Ti ). The irradiation target sample holder was a custom-made Al ladder-type frame, that held four targets in a vertical row. The outer dimensions of each target frame were 24.1 x 14.9 mm  with a 10 mm diameter collimating hole in the middle.  The Ti foils were attached to the target frames using carbon conductive cement.

\subsection{Irradiation Procedure}

Irradiation of samples was conducted in the Positive Ion Injector (PII) area of the ATLAS accelerator at the Argonne National Laboratory. Stacks of four titanium samples (total thickness 3.6 mg/cm$^{2}$) were irradiated with a 136 MeV $^{136}$Xe$^{+26}$ beam with a beam  intensity of 3 particle-nanoamperes. The beam spot size was about 0.2 cm in diameter. Three different fluences, (1.8, 5.7, and 21.5) x 10$^{14}$ ions/cm$^{2}$, were delivered using  irradiation times of 5, 15, and 60 min, respectively. The pressure in the target chamber was about 5 x 10$^{-7}$ Torr. The equilibrium temperature of the target for a beam intensity of 3 particle -nanoamperes was estimated to be 966$^{\circ}$C\cite{kantele} assuming cooling by pure radiation or pure conduction.   Before and after irradiation, the samples were examined by atomic force microscopy, scanning electron microscopy and X-ray diffraction. 

\subsection{Surface Examination}

A detailed examination of the surface of the Ti foils  was made using atomic force microscopy (VEECO/Digital Instruments). Samples were scanned using silicon nitride cantilevers in contact mode using a Multimode AFM with a Nanoscope IIIa controller and an E scanner, calibrated using a silicon grid with 10 x 10 $\mu$m$^{2}$ pitch. All images were captured at the scan rate of 1 Hz, using 256 x 256 pixel resolutions with scan sizes from 50 nm to 15 $\mu$m and flattened to remove large degrees of background tilt. The images were analyzed using the WSxM 5.0 version developed by Nanotec Electronica \cite{Horcas}. 
Scanning electron microscopy (SEM) with X-ray energy dispersive spectroscopy (EDS) (FEI Quanta 600F, Hillsboro OR, USA) was used to image the structural topologies of samples at low magnifications and to measure the elemental composition of samples. The chamber and gun pressures of SEM/EDS were operated at 1.48 x 10$^{-5}$ Pa and 1.61 x 10$^{-7}$ Pa, respectively with an emission current of 482 $\mu$A. Samples were examined at magnifications ranging between 110x and 165000x.
X-ray diffraction (XRD) data were acquired using monochromatic copper K$_{\alpha}$ radiation, a graphite diffracted beam monochromator, and scintillation detector on a Bragg-Brentano focusing diffractometer (Rigaku, Ultima IV). The X-ray tube was operated at 40 kV and 40 mA. Diffraction data was collected between two-theta angles of (0-80$^{\circ}$) with 0.02$^{\circ}$ steps with a 20 second count time per step. PDXL software from Rigaku, Inc. was used to plot the diffraction patterns, identify d-values, and fit reference patterns to identify the phase(s) present. The International Centre for Diffraction Data Ð Powder Diffraction File (ICDDPDF) database of reference patterns was used to identify phases present in samples.

\section{Results and Discussion}
\subsection{Morphology of the irradiated Ti surfaces}

When 136 MeV $^{136}$Xe$^{26+}$ ions strike the Ti foils, they create micro-structural damage due to high electronic energy loss. The formation of cavities, blisters, hillocks, dislocation loops, dislocation lines, craters and ridges as well as polygonal ridge networks has been observed by AFM. A comparison of the fluence and energy deposited in titanium samples aids in the understanding of the microstructure seen as a function of fluence. 

In Fig. 1, we show the SEM micrographs of irradiated (2.2 x 10$^{15}$ ions/cm$^{2}$) Ti foils. The SEM images (400 x) show a non-uniform distribution of spots in the irradiated Ti surface (Fig 1a). A higher magnification view (6000x) of one of the spots (which was 17$\pm$2 $\mu$m in diameter) is also shown . In another higher magnification view (165000 x) (Fig 1b) we show that the Ti has become amorphized, i.e., has been converted from a crystalline material to an amorphous material.  EDS spectra of these spots indicate only Ti is present, i.e., during or after the irradiation, the Ti was not oxidized.

\begin{figure}[H]
\centering
\includegraphics*[totalheight=3.5in]{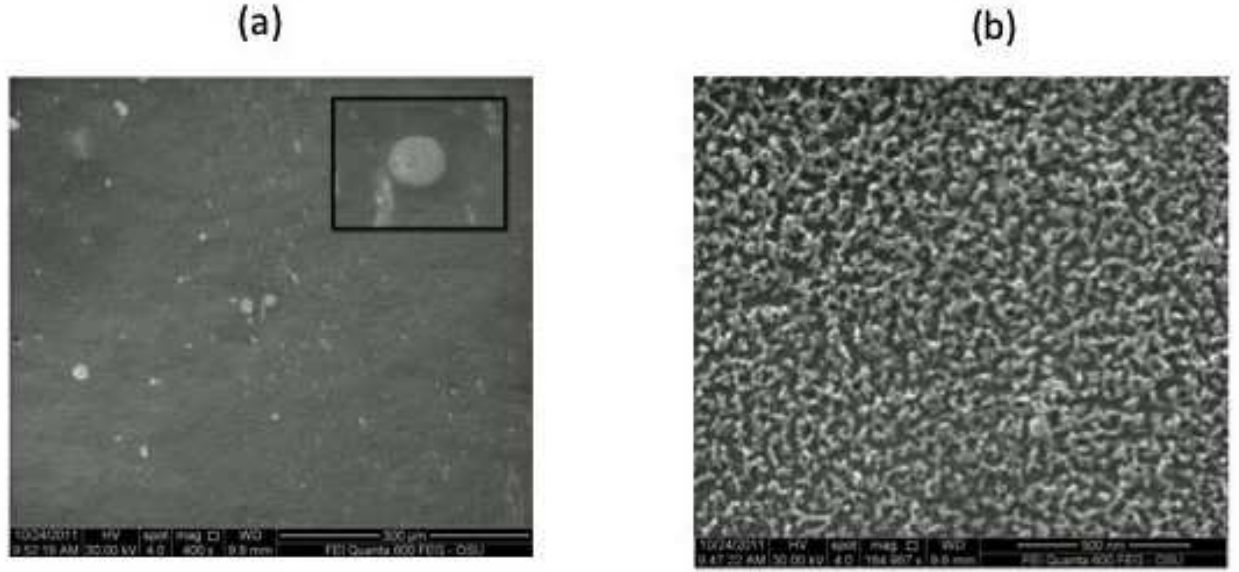}
\caption{ \label{fig1}SEM micrograph of Ti foil after irradiation with a fluence of 2.2 x 10$^{15}$ cm$^{-2}$ 136 MeV $^{136}$Xe$^{26+}$}
\end{figure}

The examination of a stack of irradiated titanium foils allowed us to study the micro-damage evolution in the electronic and nuclear stopping regimes. The AFM images of these irradiated samples at different fluences and energies are shown in Fig. 2. Figures 2(a), (e) and (i) show the microstructure damage of Ti foils irradiated with 136 MeV $^{136}$Xe$^{26+}$ with fluences of 1.8 x 10$^{14}$cm$^{-2}$, 5.4 x 10$^{14}$cm$^{-2}$ and 2.2 x 10$^{15}$cm$^{-2}$, respectively. All the scanned images correspond to 2.5 x 2.5 $\mu$m$^{2}$ areas. Through a comparison of AFM images (Fig. 2) at different fluences (1.8 x 10$^{14}$, 5.4 x 10$^{14}$ and 2.2 x 10$^{15}$ cm$^{-2}$) and ion energies (136, 93, 53, 23 MeV), we will study radiation damage as a function of the fluence and radiation damage as a function of the energy.  (The ion energies were calculated using SRIM \cite{SRIM}.)

\begin{figure}[h]
\begin{center}
\includegraphics*[scale=0.66]{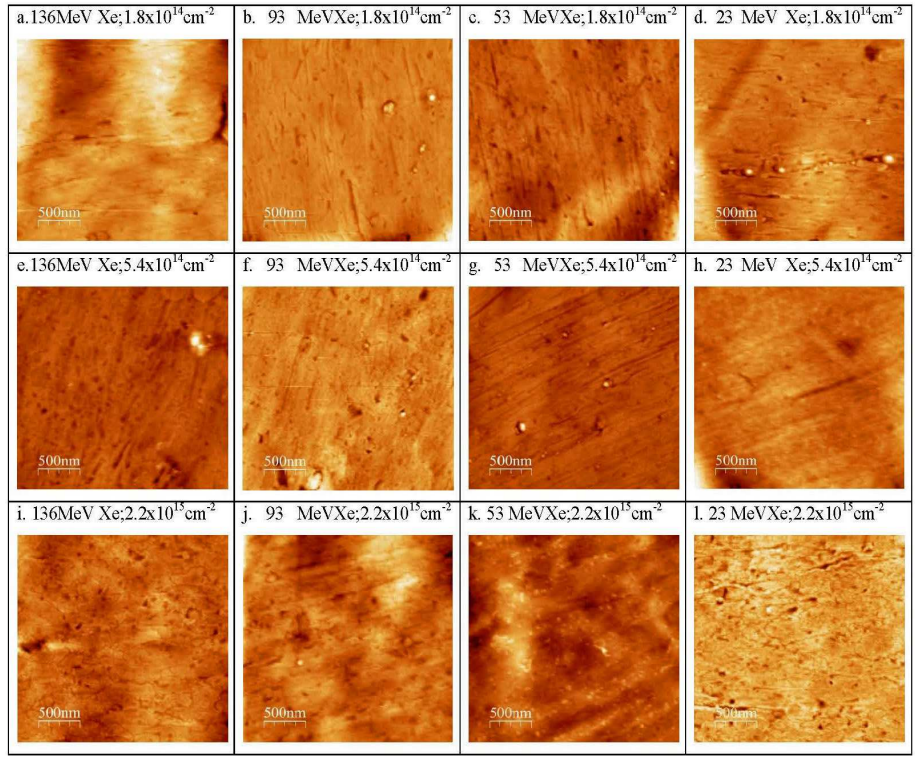}
\caption{\label{fig2}AFM images of Ti films irradiated with different fluences and energies.}
\end{center}
\end{figure}
\clearpage

\subsection{Radiation damage as a function of the fluence and energy of the incident ion.}

There are five different types of defects produced in titanium surfaces after irradiation with $^{136}$Xe$^{+26}$ ions with different fluences and energies. Those defects are hillocks, voids, polygonal ridge networks, dislocation lines and dislocation networks.

\subsubsection{Hillocks (small hills or mounds)}

Hillocks are formed when material flows out of the region of the ion track during the hot stage of the track formation process \cite{krauser}. Hillock formation is thought to be a result of interplay among surface movement of atoms and the effects of sputtering \cite{wehner}. 

During our earlier investigations, involving irradiation of Ti with 132 MeV $^{132}$Xe$^{+29}$ and fluences  up to 9.0x 10$^{13}$ ions/cm$^{2}$, hillock formation was not observed  \cite{sadi}.  When Ti foils were irradiated with 130 MeV $^{132}$Xe ions with a fluence up to 4.9 x 10$^{14}$ cm$^{-2}$, hillocks were not observed \cite{Budzynski}. However, in this work, involving higher fluences, 1.8 x 10$^{14}$ cm$^{-2}$ to 2.2 x 10$^{15}$ cm$^{-2}$, hillocks were clearly observed in all the irradiated samples. 

In Fig. 3a we show the non-irradiated titanium surface. After irradiation with 1.8 x 10$^{14}$ cm$^{-2}$ 136 MeV $^{136}$Xe$^{+26}$ (Fig 3b) hillock formation was clearly observed in the irradiated titanium surfaces. The hillock density was 8.59 $\pm$ 0.13 x 10$^{6}$ cm$^{-2}$ and the mean diameter and mean height of the hillocks were 324 $\pm$2 nm and 32 $\pm$ 1 nm, respectively. By increasing the fluence to 5.4 x 10$^{14}$ cm$^{-2}$ the density of hillocks grew from 8.59 $\pm$ 0.13 x 10$^{6}$ cm$^{-2}$ to 1.60 $\pm$ 0.27 x 10$^{7}$ cm$^{-2}$ and the mean diameter and mean height of hillocks were 364 $\pm$ 63 nm and 27 $\pm$ 2 nm, respectively. However when the fluence was increased to 2.2 x 10$^{15}$ cm$^{-2}$, the areal density, mean diameter and mean height of the hillocks decreased. (The number of hillocks created on the titanium surface was 6.81 $\pm$ 0.24 x 10$^{6}$; meanwhile, the diameter and height of hillocks were 210 $\pm$ 20 nm and 20 $\pm$ 2 nm, respectively.)   

\begin{figure}[ht]
\begin{center}
\includegraphics*[scale=0.65]{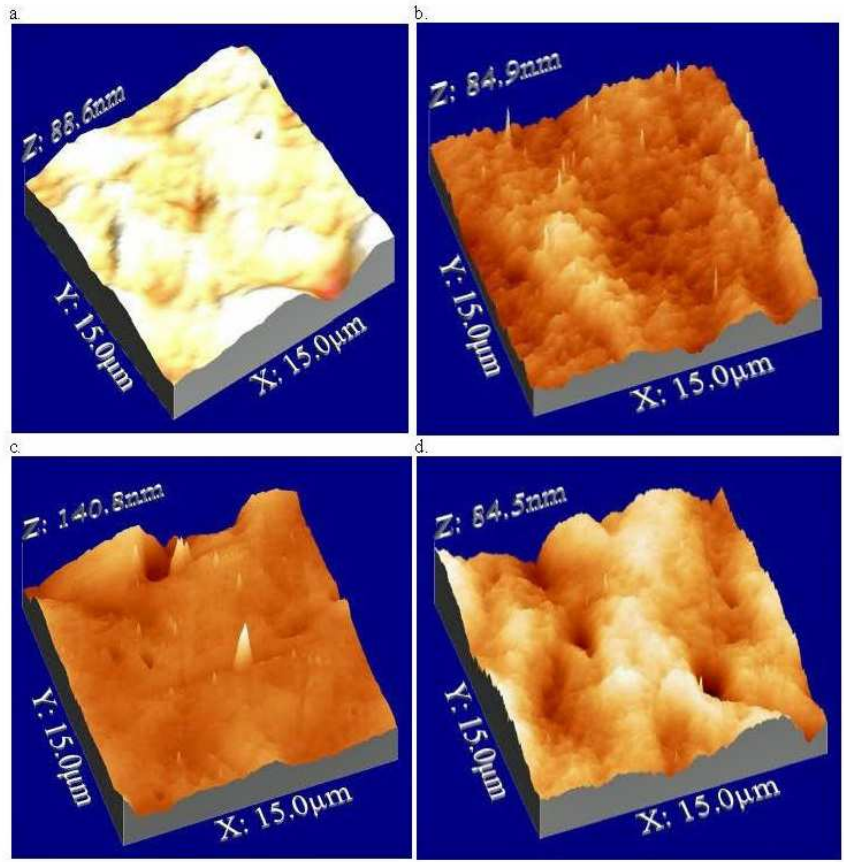}
\caption{ \label{fig 3}Three-dimensional AFM images of Ti films irradiated with 136 MeV $^{136}$Xe with fluences of 
            (a) 0, (b) 1.8 x 10$^{14}$ cm$^{-2}$, (c) 5.4 x 10$^{14}$cm$^{-2}$, (d) 2.2 x 10$^{15}$cm$^{-2}$ . 
}
\end{center}
\end{figure}
\clearpage

We can also examine the dependence of hillock formation upon the energy of the incident ions at constant fluence.  (Figure 4).  At incident ion energies of about 50  MeV, the hillock density peaks.  In Figure 5, we show a simple explanation of this effect where smaller hillocks merge to form a larger hillock with larger energy deposits.

\begin{figure}[ht]
\begin{center}
\includegraphics*[scale=0.50]{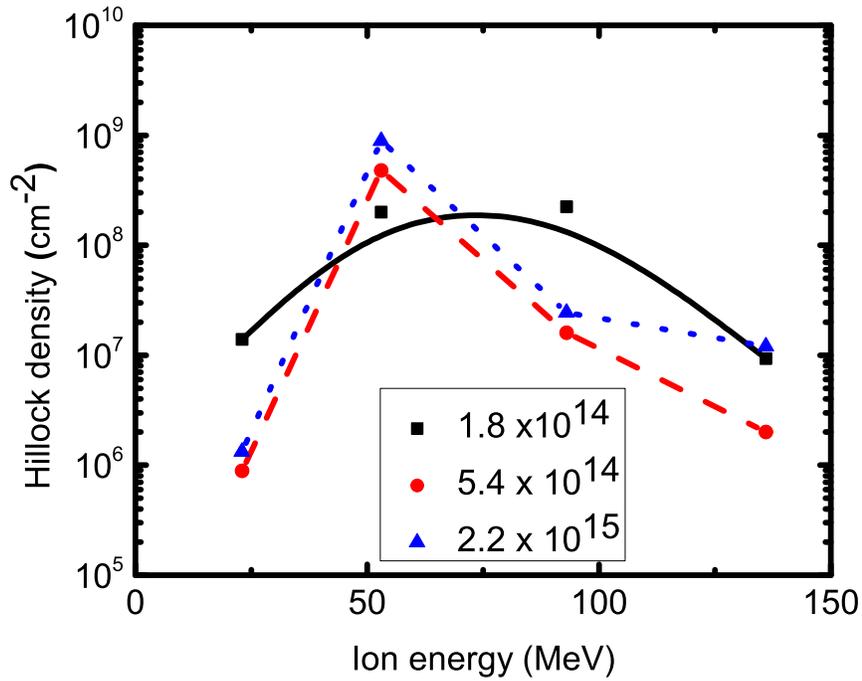}
\caption{ \label{fig 4}Energy dependence of hillock density at different fluences.}
\end{center}
\end{figure}

\begin{figure}[htb]
\begin{center}
\includegraphics*[scale=0.60]{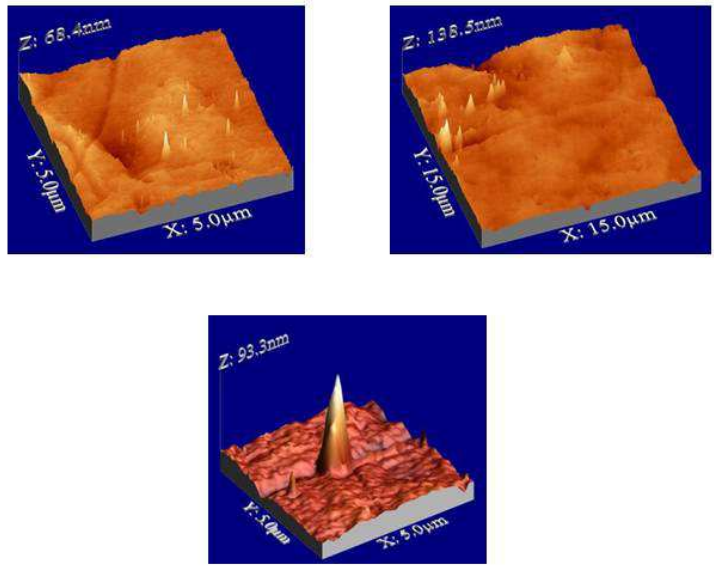}
\caption{ \label{fig 5}Coalescence of hillocks with larger energy deposits}
\end{center}
\end{figure}
\clearpage

\subsubsection{Craters/voids}

Voids/craters in the irradiated titanium foils were observed at all radiation doses. Typical AFM micrographs of void evolution in the surface of irradiated titanium foils after irradiation with 136 MeV $^{136}$Xe$^{+26}$ are shown in Fig. 2. Large doses lead to the largest void sizes. After irradiation with 1.8 x 10$^{14}$ cm$^{-2}$ ions, the voids in the irradiated Ti foils had a density of 1.18 $\pm$ 0.23 x 10$^{9}$ voids/cm$^{2}$ with a mean diameter and mean depth of 70 $\pm$ 2 nm and 5 $\pm$ 1 nm, respectively. By increasing the fluence to 5.4 x 10$^{14}$ cm$^{-2}$, the density of voids grew from 1.18 $\pm$ 0.23x10$^{9}$ voids/cm$^{2}$ to 2.72 $\pm$ 0.14 x 10$^{9}$ voids/cm$^{2}$ with the mean diameter and mean depth were 99 $\pm$ 4 nm and 7 $\pm$ 1 nm, respectively.  By increasing the fluence up to 2.2 x 10$^{15}$ cm$^{-2}$, the density of voids grew to 3.88 $\pm$ 0.48 x 10$^{9}$ voids/cm$^{2}$, although their mean sizes were not significantly increased. The growth of the voids may have been restricted by the formation of dislocation lines and dislocation networks. The dislocations function as a sink to absorb the point defects or voids.

The void density generally increased with the energy of the incident ions at constant fluence.

\subsection{Dislocations Lines, Dislocation Networks and Polygonal Ridge Networks}

After irradiation of Ti with 5.4 x 10$^{14}$ cm$^{-2}$ 136 MeV $^{136}$Xe$^{26+}$ ions, polygonal ridge networks and void-dislocation lines were observed.  (Fig 6)  The polygonal ridge networks (Fig. 6) have a density of 4.5 $\pm$ 0.5 x 10$^{6}$ polygonal ridge networks/cm$^{2}$.  The mean width and height of  the polygonal ridge networks were 303 $\pm$ 8 nm and 38 $\pm$ 5 nm, respectively. Each side of polygonal ridge networks had a length of 2.5 $\pm$ 0.5 $\mu$m. These networks are expected to lead to macroscopic increase in abrasive wear of the Ti surface.

By increasing the fluence, the density of polygonal ridge networks is expected to increase as interactions increase between projectiles and the titanium atoms. By increasing the fluence up to 2.2 x 10$^{15}$cm$^{-2}$ 136 MeV Xe ions, however, no obvious polygonal ridge networks were observed.  However, lowering the Xe energy to 53 MeV at the same fluence, resulted in the observation of a second set of polygonal ridge networks.

\begin{figure}[htb]
\begin{center}
\includegraphics*[scale=0.60]{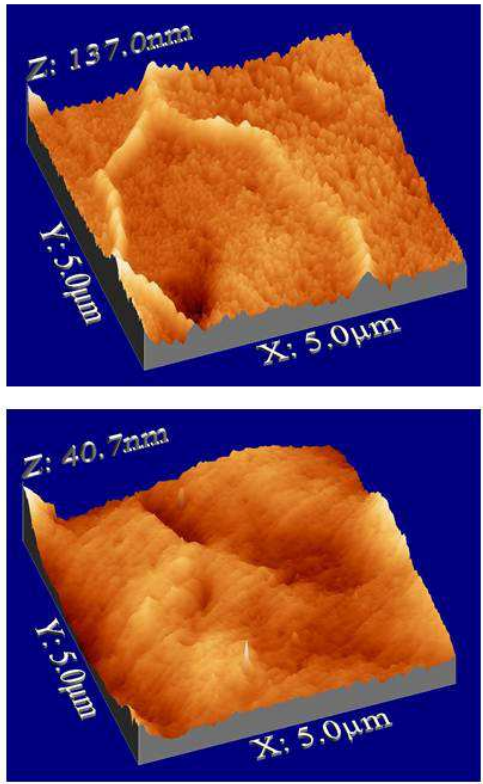}
\caption{ \label{fig 6}Three-dimensional AFM images of polygonal ridge networks (a) and void-dislocation lines (b) after irradiation with 5.4 x 10$^{14}$ cm$^{-2}$136 MeV $^{136}$Xe$^{26+}$ ions.}
\end{center}
\end{figure}
\clearpage

\subsection {Surface roughness}

In Fig. 7, we  show the root mean square (rms) surface roughness of irradiated titanium films as a function of the ion fluence. The rms surface roughness of titanium surface decreases logarithmically with increasing fluence .  Similarly the surface roughness is roughly constant as the ion energy decreases from 136 MeV until about 53 MeV. 

\begin{figure}[htb]
\begin{center}
\includegraphics*[scale=0.55]{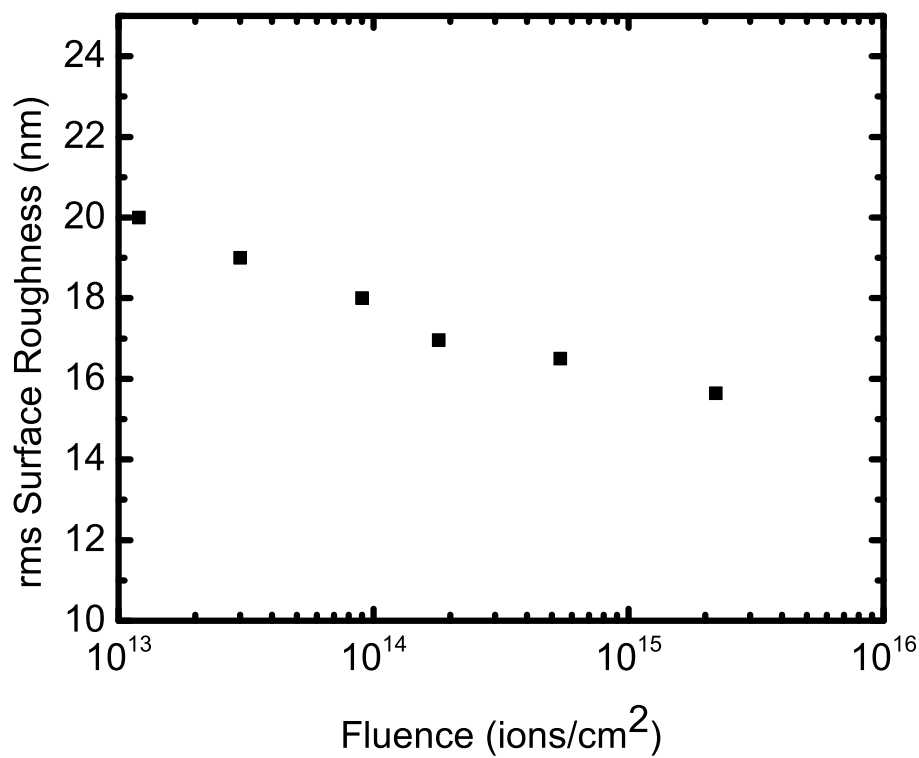}
\caption{ \label{fig 7}Ti surface roughness as a function of fluence. The un-irradiated Ti has a surface roughness of 22 nm.}
\end{center}
\end{figure}
\clearpage

\subsection{Phase stability}

In this experiment the polycrystalline titanium foil consisted of 64$\%$  fcc-Ti and 36$\%$ hcp-Ti phases before irradiation.. The measured lattice parameter of the fcc-titanium phase before irradiation was  4.0640 $\AA$; meanwhile the lattice parameters of hcp-Ti phase were a = 2.9505 $\AA$ and c= 4.6826 $\AA$ (c/a = 1.5871).  When the titanium foils were irradiated with 136 MeV $^{136}$Xe$^{+26}$ at beam intensity of 3 pnA corresponding to a foil temperature of 966 $^{\circ}$C, it was expected \cite{Lutjering} that their structure would change from hexagonal-close packed (hcp) to body-centered cubic (bcc). However, our results show that after titanium samples were irradiated with 1.8 x 10$^{14}$, 5.4 x 10$^{14}$, and 2.2 x 10$^{15}$ cm$^{-2}$ 136 MeV $^{136}$Xe$^{+26}$,  no alpha-Ti (hcp) or bcc phases were observed in the samples and merely the fcc-Ti phases were observed.  Figure 8 shows a comparison of XRD pattern of non-irradiated and irradiated Ti samples at different doses. The shrinking of (002) peak at 2$\theta$ of 38.4$^{\circ}$ in the hexagonal phase and increasing (220) peak at 2$\theta$ of 64.9$^{\circ}$ indicates that the transformation from $\alpha$-Ti (hcp) phase to fcc-Ti phase has occurred.

\begin{figure}[htb]
\begin{center}
\includegraphics*[scale=0.75]{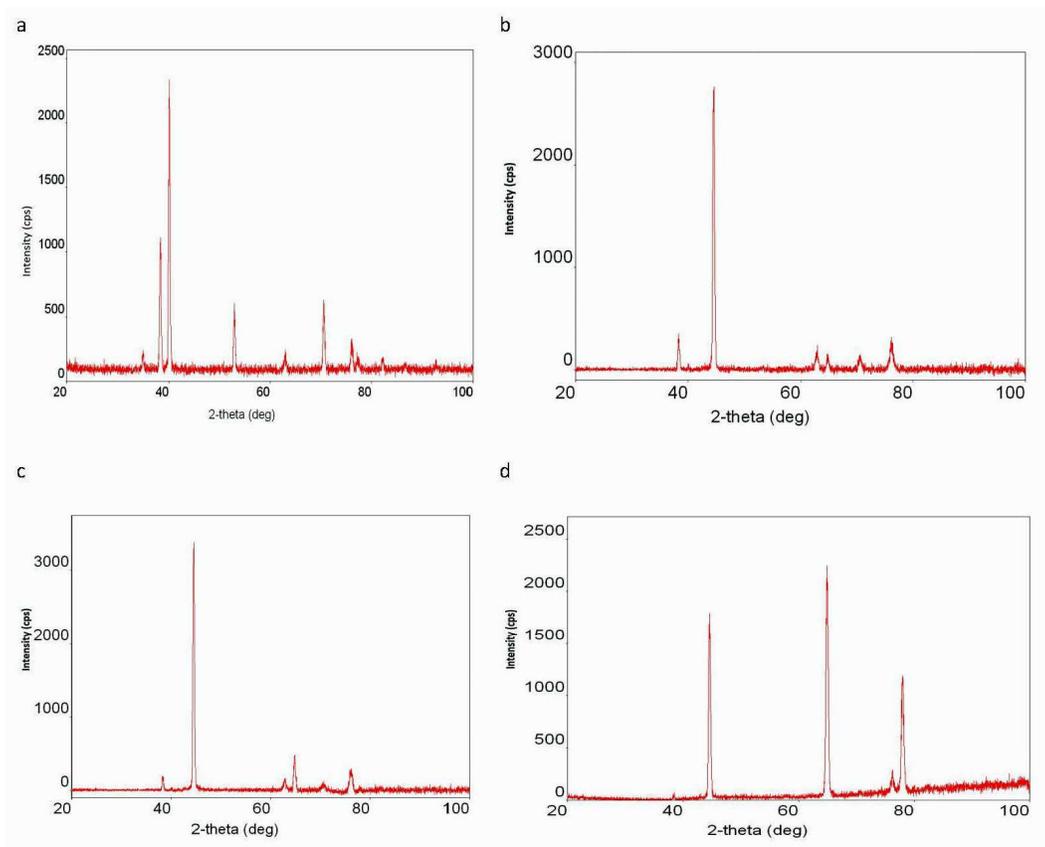}
\caption{ \label{fig 8}XRD pattern of pristine and irradiated Ti foils  at  fluences 
           of (a) 0; (b) 1.8 x 10$^{14}$; (c) 5.4 x 10$^{14}$; and (d) 2.2 x 10$^{15}$ cm$^{-2}$.   
}
\end{center}
\end{figure}
\clearpage

\subsection{Comparison of observations with SRIM simulations}

We made a SRIM \cite{SRIM} calculation to see what would be predicted for the damage due to Xe irradiation (at the highest fluence) and whether the Xe ions will create an amorphous Ti layer (see Fig. 1). 

At the damage peak (at the depth of around 10 $\mu$m), the vacancy rate is about 1.7 vacancy/target atom with assuming that 99$\%$ of the damage instantly anneals, leaving only 1$\%$ damage. SRIM calculations showed that a fluence of 2.2 x10$^{15}$ Xe ions/cm$^{2}$ of  136 MeV $^{136}$Xe$^{+26}$ ions  created 3.74 x 10$^{21}$ stable vacancies/cm$^{3}$. Since the density of titanium is 5.681 x 10$^{22}$ atoms/cm$^{3}$ and the calculated damage is 3.74 x 10$^{21}$ vacancies/cm$^{3}$, then about 6.58$\%$ of the Ti target is damaged  and the Ti layer is not amorphous. This conclusion is not quite right because the displacement energy for Ti will decrease as damage accumulates. This means that once we have partially  damaged material, it is easier to create more damage because the lattice is more loosely coupled and atoms are easier to dislodge. These changes in the crystal integrity are not included in SRIM, so the damage may be underestimated. Our result shows that the damage in term in void swelling was about 12.2$\%$ which is higher that of the SRIM calculation (6.6$\%$).

\section{Conclusions}

When a crystalline material such as titanium is subjected to 136MeV $^{136}$Xe$^{+26}$ ion bombardment at a fluence of 2.2 x 10$^{15}$ Xe ions/cm$^{2}$, lattice damage can be created ranging from point defect disorders, voids, dislocation lines, dislocation networks, polygonal ridge networks, hillocks and craters to complete lattice breakdown and new phase formation.

Phase transformation of polycrystalline titanium phase from the alpha-Ti (hcp) to the fcc phase by irradiation of titanium foils with 2.2 x 10$^{15}$ cm$^{-2}$ 136 MeV $^{136}$Xe$^{+26}$ was observed.    Phase transformation from the hcp structure  to the fcc structure  could be of considerable importance because fcc materials are generally more ductile than hcp materials.


\section{Acknowledgments}
We gratefully acknowledge the help of T. Sawyer in the SEM/EDS measurements.  This work was supported (WL) in part by the Office of High Energy and Nuclear Physics, Nuclear Physics Division, US Department of Energy, under Grant No. DE-FG06-97ER41026 and from faculty startup funds (AP) from the Department of Nuclear Engineering and Radiation Health Physics of Oregon State University.

\section{References}


\bibliographystyle{model1a-num-names}



\end{document}